\documentclass[]{spie}
\usepackage[]{graphicx}
\usepackage{amsmath}
\usepackage{epsfig}
\newcommand{\be}{\begin{equation}}
\newcommand{\ee}{\end{equation}}
\newcommand{\bea}{\begin{eqnarray}}
\newcommand{\eea}{\end{eqnarray}}
\newcommand{\ket}[1]{\left|#1\right\rangle}
\newcommand{\bra}[1]{\left\langle #1\right|}

\newcommand{\bc}{\begin{center}}
\newcommand{\ec}{\end{center}}

\newcommand{\forget}[1]{}

\newcommand{\re}{{\rm e}}
\title{Phase control of  electromagnetically induced transparency and its applications to tunable group velocity and atom localization}
\author{Kishore T. Kapale\supit{a}, Mostafa Sahrai\supit{b}, Habib Tajali\supit{b}, and  M. Suhail Zubairy\supit{c} 
\skiplinehalf
\supit{a} Quantum Computing Technologies Group, Jet Propulsion Laboratory,
California Institute of Technology, Mail Stop 126-347, 4800 Oak Grove Drive, 
Pasadena, California 91109-8099;\\
\supit{b}  Department of Physics, Tabriz University, Tabriz,
Iran; \\
\supit{c}Institute for Quantum Studies and Department of
Physics, Texas A\&M University, College Station, TX 77843-4242.
}

\authorinfo{Further author information: (Send correspondence to K.T.K. or M.S.Z.)\\ K.T.K.: E-mail: Kishor.T.Kapale@jpl.nasa.gov, Telephone: 1 818 393 7709 \\  M.S.Z.: E-mail: zubairy@physics.tamu.edu, Telephone: 1 979 862 4047}

\begin{document}
\maketitle
\begin{abstract}
We show that, by simple modifications of the usual three-level $\Lambda$-type
scheme used for obtaining electromagnetically induced transparency (EIT), 
phase dependence in the response of the atomic medium to a weak probe 
field can be introduced. This gives rise to phase dependent susceptibility. 
By properly controlling phase and amplitudes of the drive fields we obtain 
variety of interesting effects. On one hand we obtain phase control of the 
group velocity of a probe field passing through medium to the extent that 
continuous tuning of the group velocity from subluminal to superluminal and 
back is possible. While on the other hand, by choosing one of the drive fields 
to be a standing wave field inside a cavity, we obtain sub-wavelength 
localization of moving atoms passing through the cavity field. 
\end{abstract}


\section{INTRODUCTION}
\label{sec:intro}
The study of quantum coherence and quantum interference in a 
atomic system has led to a variety of  
novel phenomena such as lasing without 
inversion~\cite{1}, absorption cancellation~\cite{2}, 
refractive index enhancement~\cite{3}, electromagnetically 
induced transparency\cite{4}, ultra-slow~\cite{5}, superluminal~\cite{wang} 
and even stored light~\cite{storage},
and unprecedented control of spontaneous emission \cite{6,7,8}. 
The quantum coherence effects observed in light-matter interactions have also 
been applied to an entirely different area of atomic manipulations, namely atom
optics. There exist schemes for localization of moving atoms as they pass through 
the standing wave field of a cavity by monitoring spontaneous emission
spectrum~\cite{Zoller96,Herkomer97,QamarPRA2000,QamarOC2000,GhafoorPRA2000,GhafoorPRA2002} and by monitoring absorption of a weak probe
field~\cite{Knight2001}.

Noting that the optical fields, in general, have a carrying phase, one can 
expect that there will be phase dependence involved in atomic coherence 
effects where more than one drive fields are used due to a relative phase 
difference between them.  
Utilization of the phase dependence of atomic properties such as 
coherence have been proposed in the manipulation of the spontaneous 
emission spectra~\cite{9,GhafoorPRA2000,11} and manipulation of the index of refraction  and the group velocity of light~\cite{arbiv,SahraiGroupVel}. 
Phase dependent effects are 
fascinating due to the inherent tunability associated 
with them and they offer an easy to control parameter in an experiment.
This tunability associated with 
manipulation of relative phase allows a whole range of 
possibilities in the spontaneous emission spectrum and group velocity manipulations.  We have recently proposed a model to use phase control to obtain subluminal to superluminal light propagation in a single system~\cite{SahraiGroupVel}, and sub-half-wavelength localization of atom passing through a standing 
wave field~\cite{SahraiAL}.  It is a general understanding that one needs closed
systems to obtain phase dependence. However, as our scheme uses an open system, 
the probe field is completely independent of the conditions needed
for achieving the phase dependence.

In this article we revisit some of our recent
results~\cite{SahraiGroupVel,SahraiAL} to study the phase dependence of
susceptibility for a weak probe applied to a specially prepared four-level 
atomic medium and the applications of the scheme to tunable control of the 
group velocity and atom localization. We first discuss our model and brief discussion of the susceptibility
calculations in Sec.\ref{sec:model}. We then discuss how a continuous change
of phase gives the possibility of switching the group velocity from subluminal
to superluminal and vice versa in Sec.\ref{sec:gv}. Later we discuss how the same scheme can be used for localization of atom as it moves through a cavity in Sec.\ref{sec:al}. Finally we conclude in Sec.\ref{sec:conclude}.

\forget{Phase control is desirable as it would give a tunable know to change properties
of atomic medium by a simple experimentally controllable parameter, phase of a
drive field. In general, overeall phase of optical fields do not matter and they
do not have direct effect on physically observal properties which are real in
general.  However, there are cases in which we can engineer}

\section{The MODEL}
\label{sec:model}
\begin{figure}[ht]
\includegraphics[scale=0.45]{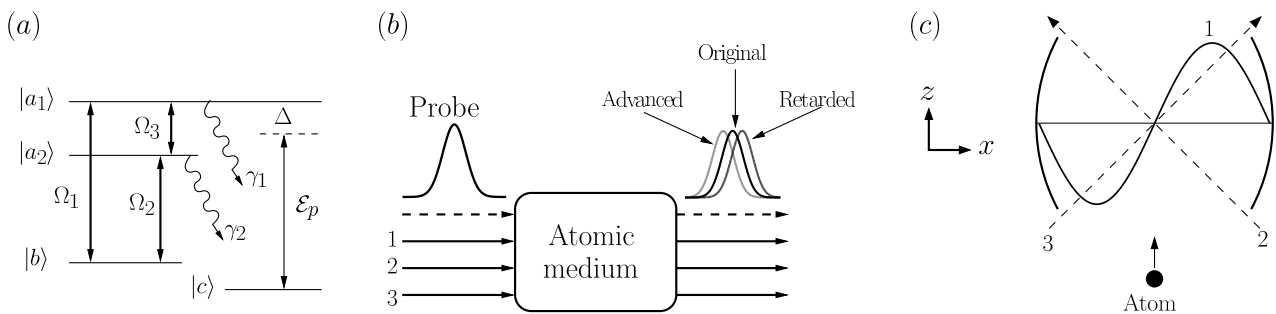}
\caption{\label{Fig:Scheme}The Model: 
$(a)$ The energy level structure
of the atom. Probe field, denoted by $\mathcal{E}_p$, is detuned by an amount
$\Delta$ from the $\ket{a_1}-\ket{c}$ transition. The labels (1, 2, 3) used 
in $(b)$ and $(c)$ part of the figure correspond to the fields with Rabi frequencies $\Omega_1$, 
$\Omega_2$ and $\Omega_3$ respectively. The decay rates from the upper levels $\ket{a_1}$ and
$\ket{a_2}$ are taken to be $\gamma_1$ and $\gamma_2$ respectively.
$(b)$
The schematic for the group velocity manipulation scheme. The three drive fields and the probe pulse all are taken to be co-propagating as they pass through the atomic medium. The relative phase of the three fields determines which one of the advanced, original or the retarded pulse will appear at the output.
$(c)$ The cavity supports the standing
wave field (1). Two other fields
(2, 3) are applied at an angle as shown. The atom enters the cavity along the $z$
axis and interacts with the three drive fields. 
The whole process takes place in the $x-z$ plane. }
\end{figure}

The schematics of the proposed scheme are shown in Fig.~\ref{Fig:Scheme}. 
Part $(a)$ of the figure shows the energy level structure of the atom as 
required by the proposal.
Schematic of the group velocity manipulation scheme is shown in part $(b)$ 
of Fig.~\ref{Fig:Scheme}. For the atom localization proposal we consider the 
setup as shown in Fig.~\ref{Fig:Scheme}(c). We
consider an atom, moving in the z direction, as it passes through a
classical standing-wave field of a cavity. The cavity is taken to be aligned 
along the x axis. The internal energy level structure of the atom is shown in 
Fig.~\ref{Fig:Scheme}(b). 

As shown in part (a) of Fig.~\ref{Fig:Scheme}, we have a four level atoms, where
the three drive fields form a complete loop.  It is this loop structure that
gives us the dependence on the relative phase difference of the three fields.
Note that the probe field is completely out of the loop and one is free to chose
the properties of the probe field as required by the experiment. Moreover, we
see that  there is a $\Lambda$ scheme ($\ket{b}-\ket{a_1}-\ket{c}$)
embedded in the level scheme, which is essential for the EIT type behavior 
that we use  for our proposal.

The Hamiltonian for the problem can be written as
\begin{equation}
\label{eq:hamiltonian1}
\mathcal{H} = \mathcal{H}_0 + \mathcal{H}_I, 
\end{equation}
where the self-energy $\mathcal{H}_0$ is given by
\begin{equation}
\label{eq:hamiltonian2}
\mathcal{H}_0 = \hbar \omega_{a_1}\ket{a_1}\bra{a_1} + \hbar
\omega_{a_2}\ket{a_2}\bra{a_2} + \hbar\omega_b\ket{b}\bra{b} +
\hbar\omega_c\ket{c}\bra{c}, 
\end{equation}
and the interaction hamiltonian is
\begin{equation}
\label{eq:hamiltonian3}
\mathcal{H}_I=-\frac{\hbar}{2} \left[ \Omega_1 \re^{- i \nu_1 t}\ket{a_1}\bra{b}
+ \Omega_2 \re^{- i \nu_2 t}\ket{a_2}\bra{b}+\Omega_3 \re^{- i \nu_3 t}\ket{a_1}\bra{a_2}
+\Omega_p \re^{- i \nu_p t}\ket{a_1}\bra{c} + \mbox{H.c.}\right].
\end{equation}
Here, $\omega_{i}$ correspond to the energy of state $\ket{i}$; the angular
frequencies of the optical fields are denoted by $\nu_{i}$; and the subscript
$p$ stands for the quantity corresponding to the probe field. 
For the rest of the discussion we assume the Rabi frequencies 
$\Omega_1, \Omega_2$ to be real and allow $\Omega_3$ to
have a carrying phase, i.e., $\Omega_3 = |\Omega_3|\re^{-i \phi}$. 
Since the three driving fields form a closed loop 
the phase can be imparted to any one of them and that will not change the result of the calculation. This will come clear at a later stage. 

The coupling
of the probe field of amplitude $\mathcal{E}_p$  is governed by the
corresponding Rabi frequency $\Omega_p=\mathcal{E}_p \wp_{a_1 c}/\hbar$. 
We note that $\wp_{a_1c}$ is the dipole moment associated with the transition
$\ket{a_1}-\ket{c}$.
We construct the density matrix equations through
\begin{equation}
\dot{\rho} = - \frac{i}{\hbar} [H,\rho]  -
\frac{1}{2} \{\Gamma,\rho\},
\end{equation}
where
$\{\Gamma,\rho\} = \Gamma\rho + \rho \Gamma$. Here the decay rates are
incorporated into the equations through the decay matrix $\Gamma$, which is
defined by 
$\langle n |\Gamma | m \rangle = \gamma_n \delta_{nm}$. 
The so obtained density matrix equations can be easily solved at steady state with the initial condition that the atom starts in its ground state $\ket{c}$ and the probe field is weak compared to all the parameters of the system. 
Solution for the important density matrix element is obtained to be 
\begin{align}
{\rho}_{a_1 c} = \frac{1}{Y \hbar}
(\Omega_2^2 - 4 \Delta^2 +  i 2 \gamma_2 \Delta) \mathcal{E}_p \wp_{a_1 c} 
\exp{(- i \nu_p t)}, 
\label{eq:tilderhoa1c}
\end{align}
where $Y=A+iB$, 
with
\begin{align}
A &= - 8 \Delta^3 +  2 \Delta (\Omega_1^2 + \Omega_2^2 +  \Omega_3^2) + 2
\gamma_1 \gamma_2 \Delta+ \Omega_1 \Omega_2 \Omega_3 (\re^{i \phi} + \re^{-i
\phi}), \nonumber \\
B &= 4 \Delta^2 (\gamma_1 +\gamma_2) - (\gamma_1 \Omega_2^2 + \gamma_2
\Omega_1^2)\,.
\end{align}
Now noting that the susceptibility can be written as
\begin{equation}
\chi = \frac{2 N \wp_{c  a_1} \rho_{a_1 c} }{ \epsilon_0
\mathcal{E}_p} \re^{i\nu_p t}
= \frac{2 N |\wp_{a_1c}|^2}{\epsilon_0} \frac{
(\Omega_2^2 - 4 \Delta^2 + i 2 \gamma_2 \Delta)}{Y \hbar},
\end{equation}
where $N$ is the atom number density in the medium.
Separating the real and imaginary parts $\chi = \chi' + i \chi''$, we obtain
\begin{align}
\chi' = \frac{2 N |\wp_{a_1 c}|^2}{ \epsilon_0 \hbar Z}
\{ (\Omega_2^2 - 4 \Delta^2) A + 2 \gamma_2 \Delta B\}\,,\qquad
\chi''  = \frac{2 N |\wp_{a_1 c}|^2}{ \epsilon_0 \hbar Z}
\{ 2 \gamma_2 \Delta A - (\Omega_2^2 - 4 \Delta^2) B)\}
\end{align}
where $Z = Y Y^*$.
It is imperative to point out that the phase enters the susceptibility expression 
only through quantities $A$ and $Y$. Moreover,
the phase dependence of $Y$ itself is only through that of $A$. 
We observe that the phase dependent term in $A$ is 
$\Omega_1 \Omega_2 \Omega_3 (\re^{i \phi} + \re^{-i\phi})$. Thus, 
this phase factor could very well have come from either of the three 
driving fields. Moreover, in the event that all the fields have a carrying 
phase, only the collective phase would occur in the susceptibility and
there would be no dependence on individual phases. 
This collective phase can be easily determined to be 
$\phi=\phi_2+\phi_3-\phi_1$, by repeating the susceptibility calculation
and noting that the Rabi frequencies are in general complex. 
Here $\phi_i$ is the phase of 
the complex Rabi frequency $\Omega_i$ of the $i$th driving field. 

Noting the phase denendence of the atomic response to a weak probe field 
it can be seen that variety of effects are possible, we study two examples 
in the following sections.

\section{PHASE CONTROL OF THE GROUP  VELOCITY}
\label{sec:gv}

For the discussion of this section the relevant quantity to consider is the group index
$n_{g} = c/v_{g}$, where $c$ is the speed of light in vacuum and the 
group velocity $v_g$ is given by
\begin{equation}
v_g=\frac{c}{1 + 2 \pi \chi'(\nu_p) + 2 \pi \nu_p \partial
\chi'(\nu_p)/\partial \nu_p}\,.
\end{equation}
Note that the group velocity depends only on the real part of the susceptibility
$\chi'$. However, the real and imaginary parts of the susceptibility are related
with each other through the Kramer-Kr\"{o}nig relations, thus careful
considerations of both the quantities are needed to find proper consitions for
obtaining phase control of the group velocity.

\begin{figure}[ht]
\centerline{\includegraphics[scale=0.385]{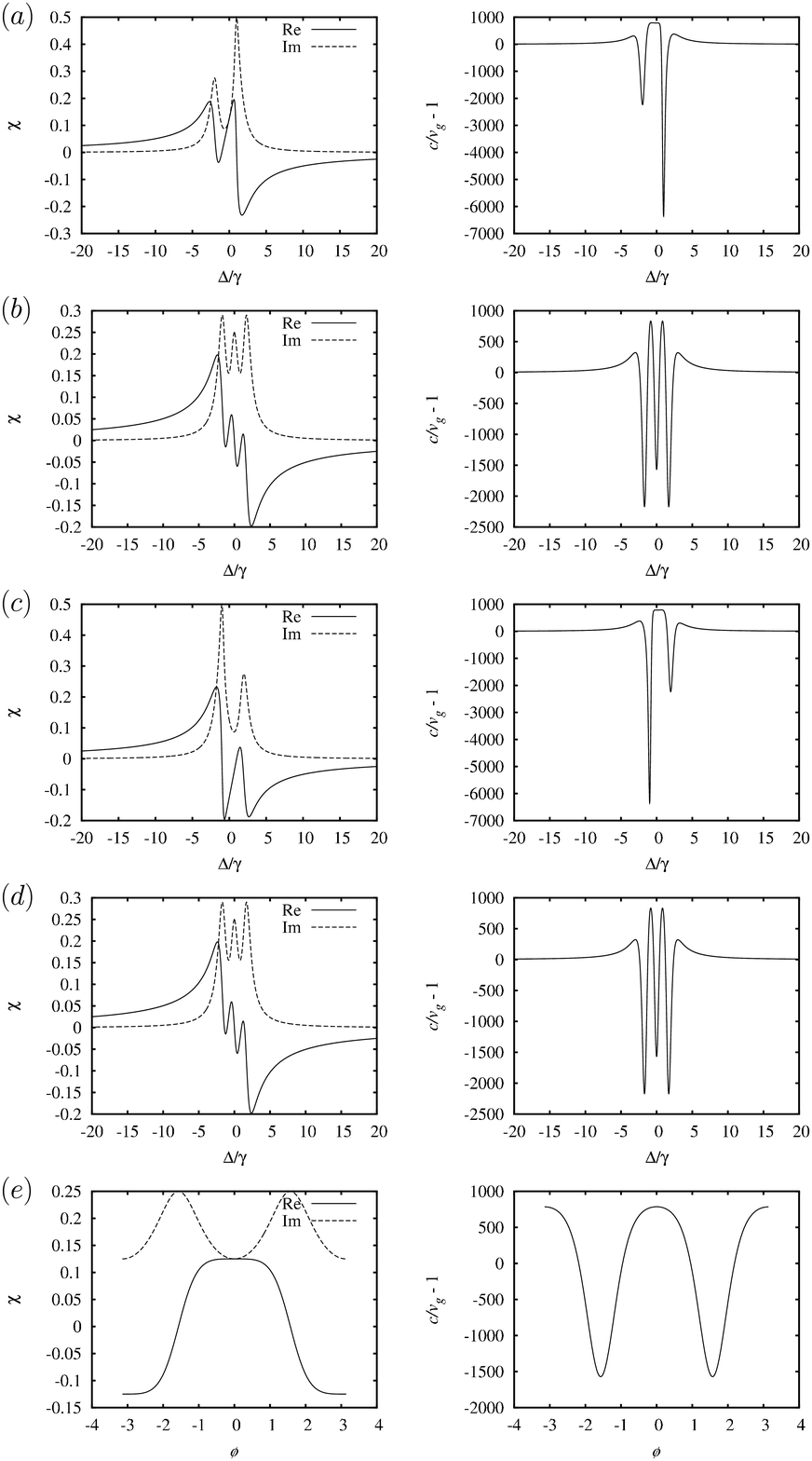}}
\caption{\label{FIG:HAbsPhaseT} Phase variation of Group Velocity 
(Accompanied by absorption): The general parameters are
$\Omega_1=\Omega_2=\Omega_3= 2\gamma$, $\gamma_1=\gamma_2=2\gamma$.
(a)$\phi=0$, (b) $\phi=\pi/2$, (c) $\phi=\pi$, (d) $\phi=3\pi/2$, and (e)
Variation of group index as a function of $\phi$.
Units: $\chi$ and $c/v_g-1$ are in the units of 
$2 N |\wp_{a_1 c}|^2/\epsilon_0 \hbar$, $\Delta/\gamma$ is dimensionless and 
$\phi$ is in radians. Please note that each label corresponds to both the plots appearing horizontally right to next to it.
}
\end{figure}

\begin{figure}[ht]
\centerline{\includegraphics[scale=0.36]{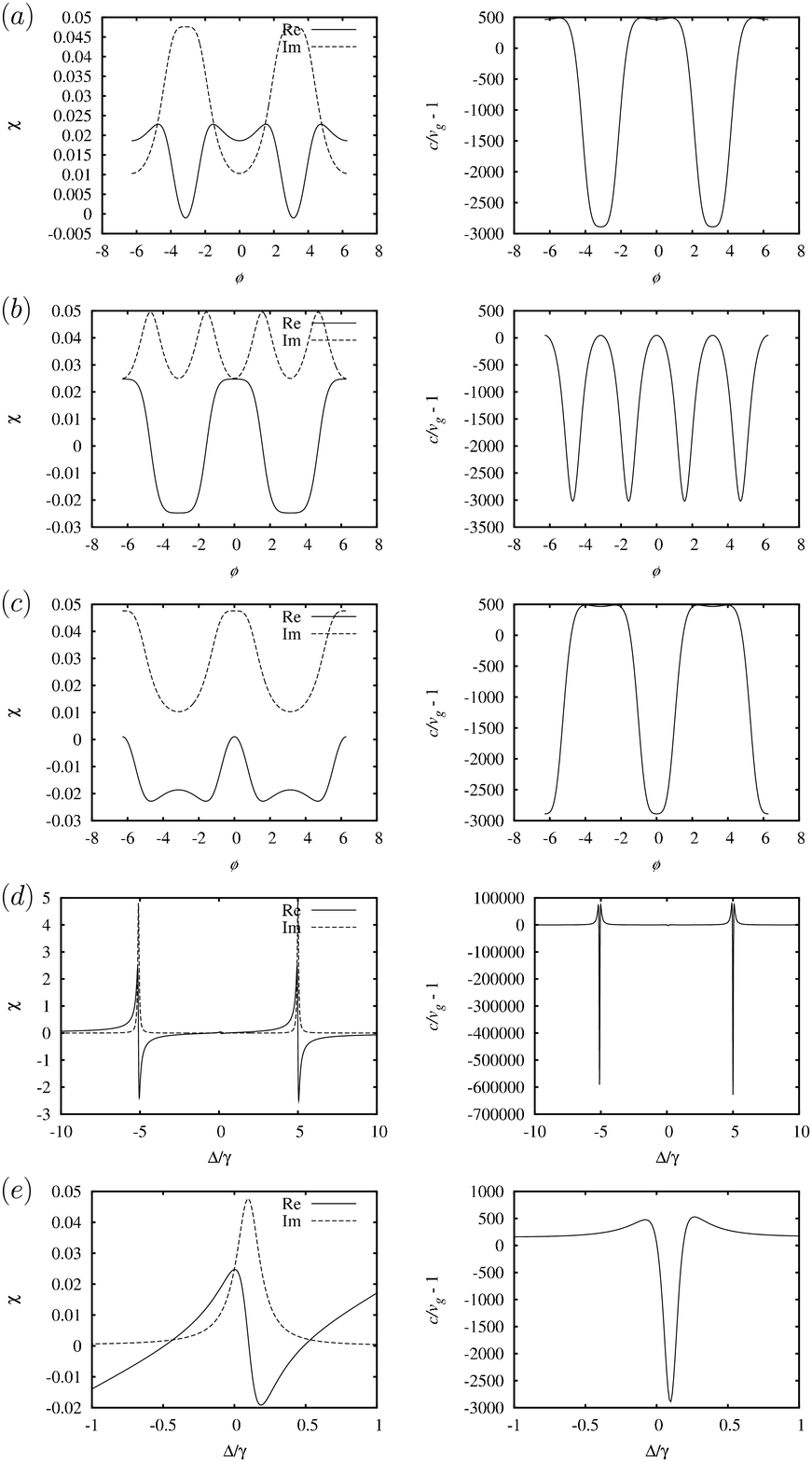}}
\caption{\label{FIG:LAbsPhaseT}Phase variation of group velocity 
(Small absorption): The general parameters are
$\Omega_1=10\gamma,\Omega_2=\Omega_3= \gamma$, $\gamma_1=\gamma_2=0.2\gamma$.
(a)$\Delta=0.1\gamma$, (b) $\Delta=0\gamma$, (c) $\Delta=-0.1\gamma$, (d)
Variation as a function of $\Delta/\gamma$ for $\phi=0$. We concentrate on the
small feature near $\Delta/\gamma=0$ in the susceptibility curves to obtain
(a)-(c). This small feature seen in (d) is magnified and depicted in (e). Thus have superluminality at a considerably less absorption compared to
most of the existing proposals. Another point to be noted here is that large
variation of group velocity is available for $\Delta/\gamma\neq 0$ but close to
$0$. The units are the same as in Fig.~2.
}
\end{figure}

We observe that our model imparts unprecedented control over the group velocity
of the probe pulse. The group velocity shows continuous tunability over a wide
range of values ranging from subluminal to superluminal with just the change of
the phase of one of the control fields while other parameters are kept constant.
Important feature of our model being considerably less absorption accompanying
the superluminal group  velocities. Except for the first experimental
realization~\cite{wang} where superluminality was gain assisted, most of the
other proposals observe superluminality along with considerable absorption of
the pulse as it passes through a specially prepared medium.
We discuss our results with the help of the figures~\ref{FIG:HAbsPhaseT}, 
\ref{FIG:LAbsPhaseT}. 

In Fig.~\ref{FIG:HAbsPhaseT} (a)-(d) we plot the susceptibility ($\chi$) and the
group index ($c/v_g -1$) as we vary the phase ($\phi$) of the field
corresponding to the Rabi frequency $\Omega_3$. We observe appearance and
disappearance of a peak in the absorption profile in the center near
($\Delta/\gamma = 0$). Thus there is  resulting change in the sign of the slope of the
dispersion  with the phase giving rise to switch in the group velocity of the
probe pulse from subluminal to superluminal. This change is continuous as
depicted in Fig.\ref{FIG:HAbsPhaseT} (e). In this choice of parameters one has
superluminality accompanied by slight absorption. Note that we have chosen a
modest value of the frequency of the probe field $\nu_p=1000\gamma$. The derivative 
term being the dominating one in the group index, for real
experimental parameters we expect the range of variation of the group velocity
to be much larger than shown here. We note that in all the figures to follow,
susceptibility $\chi$ and the group index $n_g-1$ are
plotted in the units of $2 N |\wp_{a_1 c}|^2/\epsilon_0 \hbar$. Also, 
detuning $\Delta$ is represented by its dimensionless equivalent 
$\Delta/\gamma$ and the phase $\phi$ is in radians.

However, it would be desirable to have superluminal propagation with reduced
absorption to make sure that the pulse is not attenuated considerably as it
passes through the medium. We consider a different range of parameters in 
Fig.~\ref{FIG:LAbsPhaseT} that gives us essentially the same feature. To
illustrate,  we observe that the susceptibility has a small feature near the
line center ($\Delta/\gamma=0$) as shown in part (d) of the figure. We
concentrate on this small feature and take advantage of the fact that the
absorption is small to show phase tuning of the group velocity in parts (a)-(c).
It turns out that having the probe field slightly detuned from the probe
transition gives a wider range of group velocities (parts (a) and (c)) compared
to zero detuning (part (b)). It is also instructive to notice that even though
we are focusing on a small spectral feature of the medium the range of the
group velocities available is no less than that observed in
Fig.~\ref{FIG:HAbsPhaseT}.

Now we contrast some of the features of the above considered model with some other approaches. A desirable feature in the group velocity control is intensity dependent
tunability, demonstrated by Agarwal {\it et al.}~\cite{agarwal}. Our model has this feature as well and shows the intensity controlled tuning of the group velocity~\cite{SahraiGroupVel}.  Most of the proposals for the tunability of the group 
velocity are affected adversely due to the Doppler broadening of the medium.  As pointed out by Goren {\it et al.}~\cite{goren} the tunability from subluminal to superluminal is lost completely due to the Doppler broadening effects in some cases. It is however instructive to note that the model we propose here is naturally Doppler free if we consider all the drive fields and the probe field to be propagating colinearly. To put this in perspective we note that the major requirements of our model constitute  maintaining the loop formed by the three driving fields and the two-photon resonance condition of EIT among the driving field $\Omega_1$ and the probe field. We note that since $\nu_1 = \nu_2 + \nu_3$, the Doppler shifts for collinear propagation would satisfy $\Delta \nu_1 = \Delta \nu_2 + \Delta \nu_3$ maintaining the loop structure and phase sensitivity. With the assumption that the low lying levels $\ket{b}$ and $\ket{c}$ are very close to each other the frequency shift in the driving field corresponding to $\Omega_1$ and the probe field are nearly the same thus maintaining the two-photon resonance with the transition $\ket{b}-\ket{c}$. Within the two-photon resonance regime, the only requirement for keeping the EIT medium Doppler free is to have sufficiently strong driving field~\cite{matsko}. Thus, noting that we work very close to the EIT condition our model is naturally Doppler free and there is very little absorption of any of the fields shining on the medium. Moreover, the model
maintains the phase sensitivity as it preserves the loop structure even with the Doppler shifts.

In the next section we present another application of the phase dependence of
the susceptibility of a weak probe field for atom localization.

\section{SUB-HALF-WAVELENGTH LOCALIZATION}
\label{sec:al}
For the  purpose of the discussion for this section we concern ourselves only
with the imaginary part of the susceptibility which is given as
\begin{align}
\chi'' & =
\frac{2 N |\wp_{a_1 c}|^2 }{ \epsilon_0 \hbar Z} \{ 2 \gamma_2
\Delta A - (\Omega_2^2 - 4 \Delta^2) B)\},
\label{Eq:Chiprime2}
\end{align}
However, the quantities $A$ and $B$ are modified because the field $1$ now
corresponds to the standing-wave field of the cavity as shown in
Fig.~\ref{Fig:Scheme}(c).  Therefore  the Rabi frequency $\Omega_1$ is to be 
replaced by $\Omega_1(x) = \Omega_1 \sin{\kappa x}$. Thus the new definitions of
the quantities $A$ and $B$ are given by
\begin{align}
A &= - 8 \Delta^3 +  2 \Delta \,(\Omega_1^2 \sin^2 \kappa x + \Omega_2^2 +
\Omega_3^2  ) + 2 \gamma_1 \gamma_2 \Delta +
\Omega_1 \Omega_2 \Omega_3 (e^{i \phi} + e^{-i \phi}) \sin
\kappa x , \nonumber \\ B &= 4 \Delta^2 (\gamma_1 +\gamma_2) - 
\gamma_1\, \Omega_2^2 - \gamma_2\, \Omega_1^2 \sin^2 \kappa x\,,
\end{align}
and $Z=A^2 + B^2$.
In the next section we consider the imaginary part of the susceptibility
$\chi''$in detail and obtain various conditions for sub-wavelength localization of the atom.

We study the expression~\eqref{Eq:Chiprime2} for the imaginary part of the susceptibility on the probe transition in greater details in the following 
discussion. It is clear that $\chi''$, i.e., probe absorption depends 
on the controllable parameters of the system  like probe field detuning,
amplitudes and phases of the driving fields. 

Noting the dependence of $\chi''$ on $\sin{\kappa x}$, we can see that it  is, in principle, possible to  obtain information
about the $x-$position of the atom as it passes through the cavity 
by measuring the probe absorption. Nevertheless, for precise localization of the
atom the susceptibility should show maxima or peaks along the
$x-$coordinate. We obtain the conditions for the presence of peaks in $\chi''$
in the discussion to follow. In the case of $\gamma_{2}=0$, 
i.e., the level $\ket{a_{2}}$ is a metastable  Eq.~\eqref{Eq:Chiprime2} can be simplified as follows:
\begin{align}
\chi'' &= \frac{2 N |\wp_{a_1c}|^2}{\hbar \epsilon_0} 
\frac{\gamma_1 (\Omega_2^2 - 4 \Delta^2)^2}
{\gamma_1^2(\Omega_2^2 -4\Delta^2 )^2 + ( 8 \Delta^3 
- 2\,\Delta \,(\Omega_1^2\,\sin^2{\kappa x} + \Omega_2^2 - \Omega_3^2)
-  \Omega_1 \Omega_2 \Omega_3\, \cos\phi\, \sin{\kappa x})^2}
\nonumber \\
&=\frac{2 N |\wp_{a_1c}|^2}{\hbar \epsilon_0} 
\frac{\gamma_1 (\Omega_2^2 - 4 \Delta^2)^2}
{\gamma_1^2(\Omega_2^2-4\Delta^2 )^2 
+ (\sin{\kappa x}-R_1)^2(\sin{\kappa x}-R_2)^2}
\label{Eq:chiroot}
\end{align}
where
\begin{equation}
R_{1,2}=\frac{1}{4
\Delta \Omega_1}\left( {-\Omega_2 \Omega_3 \cos\phi}
\pm \sqrt{\Omega_2^2\, \Omega_3^2\, \cos^2{\phi} - 16 \Delta^2 [ (
\Omega_2^2 + \Omega_3^2 )- 4 \Delta^2 ]}\right)\,.
\end{equation}
Thus peaks will occur in  $\chi''$ at $x-$positions satisfying 
$\sin{\kappa x } = R_{1,2}$. In other words, $\chi''$ peaks 
at the spatial positions defined by $\kappa x = \sin^{-1}(R_{1,2}) \pm n\pi$,
\forget{\begin{align}
\kappa x = \sin^{-1} \Bigl[ \left.\frac{1}{4
\Delta \Omega_1}\right( {-\Omega_2 \Omega_3 \cos\phi}
\pm \left.\sqrt{
\Omega_2^2\, \Omega_3^2\, \cos^2{\phi} - 16 \Delta^2 [ (
\Omega_2^2 + \Omega_3^2 )- 4 \Delta^2 ]}\right)\Bigr]
\pm n \pi,
\end{align}}
where $n$ is an integer, leading to localization of atoms conditioned on the 
detection of the probe absorption at that particular frequency corresponding 
to the value of $\Delta$. In general, there will be four distinct
peaks spread over the whole wavelength as both the roots 
$R_{1,2}$ appear twice in the denominator of Eq.~(\ref{Eq:chiroot}).

\begin{figure}[ht]
\includegraphics[scale=0.4]{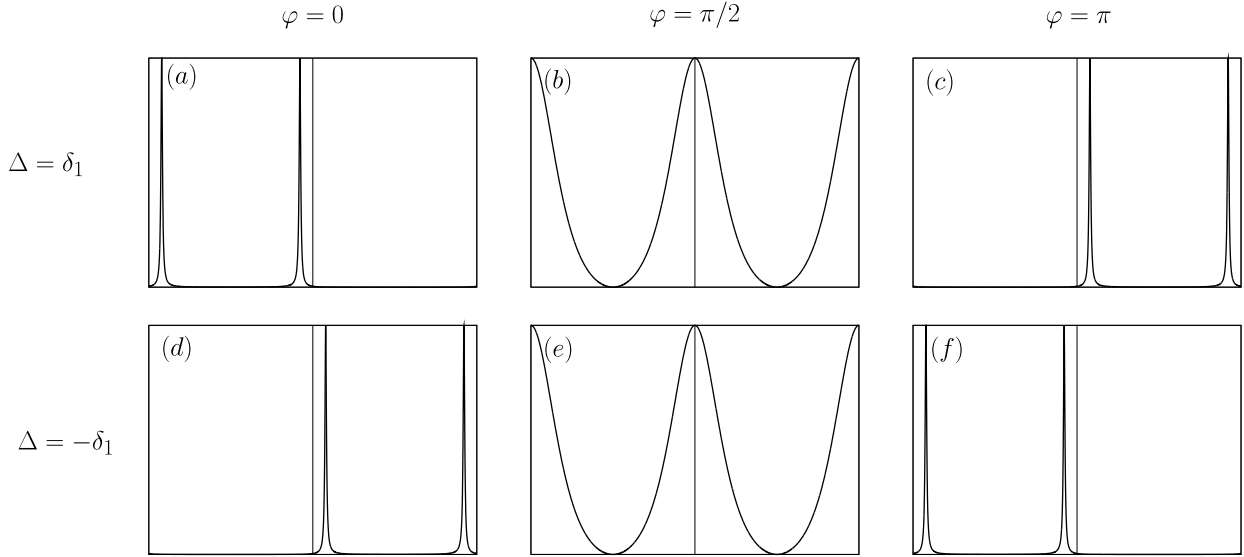}
\caption{\label{Fig:plots2} Phase dependence of the localization for the 
choice of the probe detuning  $\Delta=\pm \delta_1$ defined in
Eq.\eqref{Eq:delta12}: Plot of the imaginary part of the susceptibility in
arbitrary units vs the dimensionless $x-$coordinate $\kappa x$ along the standing
wave in the cavity.  $\kappa x$ runs from the values $-\pi$ on the extreme left
to $\pi$ to the extreme right in each box. A vertical line is drawn at $\kappa x
= 0$ for the guiding of the eye. The common parameters are 
$\Omega_2= \Omega_3=\Omega=20\gamma_1$, $\gamma_2=0$, and $\Omega_1 = 30\gamma_1$, and
$\delta_1= (\Omega/4)(\sqrt{3}-1)$ unless specified otherwise.
(a) $\phi=0$ and  $\Delta=\delta_1$
(b) $\phi=\pi/2$ and $\Delta=\delta_1$
(c) $\phi=\pi$ and $\Delta= \delta_1$
(d) $\phi=0$ and $\Delta=-\delta_1$
(e) $\phi=\pi/2$  and $\Delta=-\delta_1$
(f) $\phi=\pi$ and $\Delta=-\delta_1$.
Notice the presence of sub-half-wavelength localization for the boxes (a), (c), (d) and (f), where the two peaks are confined to either the range $\kappa x=\{-\pi,0\}$ 
or $\kappa x=\{0,\pi\}$. The range in which the localization peaks appear
depends on the value of $\phi$ and the sign of the detuning $\Delta$. 
There is no sub-half-wavelength localization for boxes (b) and (e) as 
$\phi=\pi/2$. Notice that the lineshapes are much sharper for the case of $\phi=0,\pi$ compared to $\phi=\pi/2$.}
\end{figure}

\begin{figure}[ht]
\includegraphics[scale=0.4]{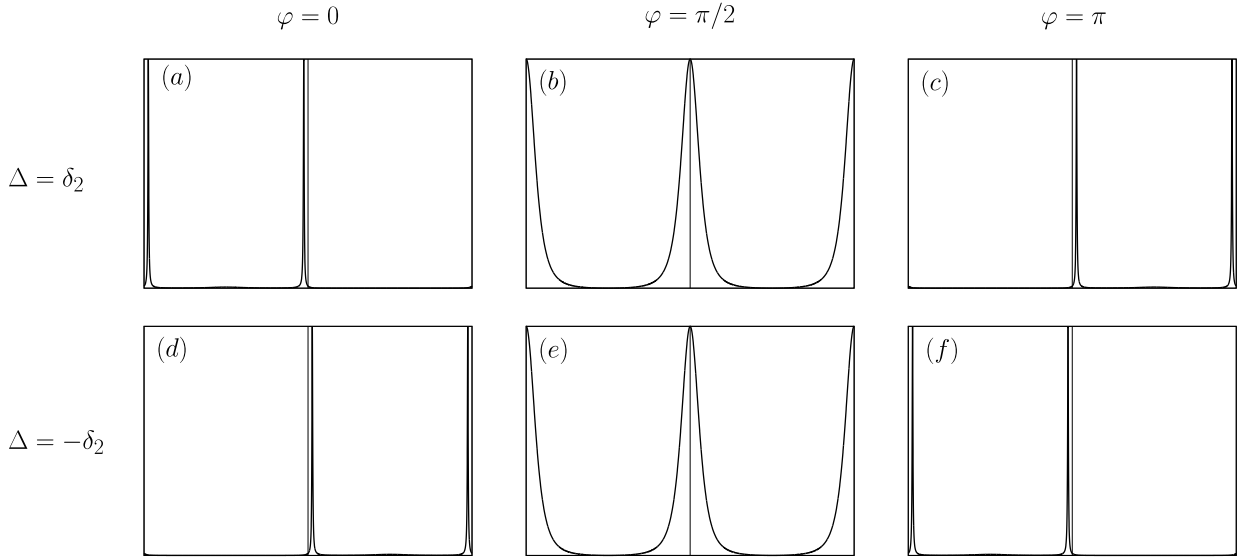}
\caption{\label{Fig:plots3} Phase dependence of the localization for the 
choice of the probe detuning $\Delta=\pm \delta_2$ defined in
Eq.\eqref{Eq:delta12}: Plot of the imaginary part of the susceptibility in
arbitrary units vs the dimensionless $x-$coordinate $\kappa x$ along the standing
wave in the cavity.   $\kappa x$ runs from the values $-\pi$ on the extreme left
to $\pi$ to the extreme right in each box. A vertical line is drawn at $\kappa x
= 0$ for the guiding of the eye. The common parameters are 
$\Omega_2= \Omega_3=\Omega=50\gamma_1$, $\gamma_2=0$, and $\Omega_1 = 60\gamma_1$, and 
$\delta_2= (\Omega/4)(\sqrt{3}+1)$ unless specified otherwise.
(a) $\phi=0$ and  $\Delta=\delta_2$
(b) $\phi=\pi/2$ and $\Delta=\delta_2$
(c) $\phi=\pi$ and $\Delta= \delta_2$
(d) $\phi=0$ and $\Delta=-\delta_2$
(e) $\phi=\pi/2$  and $\Delta=-\delta_2$
(f) $\phi=\pi$ and $\Delta=-\delta_2$. Notice the presence of sub-wavelength localization for the boxes (a), (c), (d) and (f), where
the two peaks are confined to either the range $\kappa x=\{-\pi,0\}$ 
or $\kappa x=\{0,\pi\}$. The range in which the localization peaks appear
depends on the value of $\phi$ and the sign of the detuning $\Delta$. 
There is no localization for boxes (b) and (e) as $\phi=\pi/2$. 
Notice that the lineshapes are much sharper for the case of $\phi=0,\pi$ compared to $\phi=\pi/2$. Also notice that the lineshapes in this figure are sharper as compared to the ones in  Fig.~\ref{Fig:plots2} due to larger value of $\Omega_{1}$.}
\end{figure}

As already discussed, the positions of maxima are strongly dependent on the probe 
field frequency through its detuning $\Delta$. One of the  
main requirement for obtaining the novel result we mentioned in the 
introduction, namely  sub-half-wavelength localization of the atom, 
would be to reduce the usual four peaks to 
two and to confine them to one of the half-wavelength regions along t
he standing-wave of the cavity. Once again observing  
expression~(\ref{Eq:chiroot}) for the probe absorption we can see that 
two peaks are possible only if $R_{1}$ and $R_{2}$ coincide. 
This gives rise to the condition
$
\Omega_2^2\, \Omega_3^2\, \cos^2{\phi} - 16 \Delta^2 [ (
\Omega_2^2 + \Omega_3^2 )- 4 \Delta^2 ]=0
$.
To simplify the discussion we consider a special case when $\Omega_{2}=\Omega_{3}=\Omega$. This assumption is general enough and easy to realize as the drive field strengths are controllable.
This condition can be solved for $\Delta$ with $\phi=0, \mbox{ or }, \pi $  
to obtain
\begin{align}
\Delta=\pm \frac{\Omega}{4}(\sqrt{3}\mp 1) = \pm \delta_{1,2}\, .
\label{Eq:delta12}
\end{align}
For brevity we have introduced the notation, 
$\delta_{1,2}=({\Omega}/{4})(\sqrt{3}\mp1)$, for the interesting values of the 
detunings.  We plot the  susceptibility in arbitrary units versus the 
$\kappa x$ in Figs.~\ref{Fig:plots2} and~\ref{Fig:plots3} for the detunings 
$\Delta=\pm\delta_{1}$ and $\Delta=\pm\delta_{2}$ respectively for the 
different values of the cavity field phase $\phi=0,\pi/2, \mbox{ and } \pi$ 
for the choice of the drive field Rabi frequencies such that the localization 
peaks are sharp and well within a half-wavelength when sub-localization occurs.

Another interesting value of the detuning should be looked at as well 
corresponding to the phase value $\phi=\pi/2$, to check if  there is a 
possibility of reducing the number of peaks from four to two in this case as 
well.  In fact, it can be seen that one can obtain $R_{1}=R_{2}$ for 
$\phi=\pi/2$ by solving,
$
- 16 \Delta^2 [ (
\Omega_2^2 + \Omega_3^2 )- 4 \Delta^2 ]=0
$ for $\Delta$. Thus, with $\Omega_{2}=\Omega_{3}=\Omega$ we obtain
$
\Delta=0, \text{ and } \Delta=\pm{\Omega}/{\sqrt{2}}=\pm\delta_{3}\,.
$
It can be easily seen that for these values of $\Delta=\pm\delta_{3},0$
there is no sub-half-wavelength localization possible. Now we discuss the results presented in Figs.~\ref{Fig:plots2}-\ref{Fig:plots3}
one by one.

We observe in Fig.~\ref{Fig:plots2}, where we consider 
$\Delta=\omega_{a_{1 c}} - \nu_{p}=\pm\delta_{1}$, 
that for the values of the phase of the cavity field $\phi=0$ 
(boxes $(a)$, $(d)$) and $\phi=\pi$ (boxes $(c)$ and $(f)$) there 
is sub-half-wavelength localization possible.  It can be seen that for a 
given value of the detuning changing the phase $\phi$ from $0$ to $\pi$ 
shifts the localization peaks from one-half of the wavelength of the cavity 
field to the other (See boxes $(a)$ and $(c)$ for $\Delta=+\delta_{1}$ and 
boxes $(d)$ and $(f)$ for detuning $\Delta=-\delta_{1}$ Similarly, for a 
given value of the phase $\phi$ changing the sign of the detuning moves 
the localization peaks from one half-wavelength to the other 
(See boxes $(a)$ and $(d)$ for $\phi=0$ and boxes $(c)$ and $(f)$ 
for $\phi=\pi$). Exactly similar observation applies to 
Fig.~\ref{Fig:plots3} where we have chosen $\Delta=\pm\delta_{2}$. 
We notice that $\delta_{1} < \delta_{2}$, thus different values for the 
Rabi frequencies are needed for the drive fields $\Omega_{1}$ and 
$\Omega_{2}$ to obtain well localization of the atom in a half-wavelength. 
This will be clear from the discussion to follow.

To clarify conditions to obtain good localization peaks well-within the half-wavelength region of the cavity field we remind ourselves us that the sub-half-wavelength localization occurs when $R_{1}=R_{2}$, i.e. when 
$
\sin{\kappa x} = \pm {\Omega_{2}\Omega_{3}}/({4 \Delta \Omega_{1}})=\pm \sin {X_{0}}
$
with $\phi= 0, \mbox{ or } \pi$. Note that the $+$ sign corresponds to $\phi=0$ and $-$ corresponds to $\phi=\pi$. Thus the base peak occurs either the left or right side of $\kappa x = 0$ depending on the value of the phase $\phi$. We notice that $X_{0}$ should be sufficiently far from zero so that the whole lineshape
is contained  within the given half-wavelength region. This observation governs the choice of the intensities or the Rabi frequencies of the drive fields at the chosen detuning $\Delta$. We have chosen the appropriate values for both the figures~\ref{Fig:plots2} and~\ref{Fig:plots3} to obtain sub-half-wavelength localization. It is to be noticed that for the choice of parameters corresponding to sub-half-wavelength localization for $\phi=0,\pi$,  changing the phase to $\phi=\pi/2$, overall we still obtain two peaks. However these peaks are distributed over the full wavelength as can be seen from boxes $(b)$ and $(e)$ of both the figures~\ref{Fig:plots2} and~\ref{Fig:plots3}. There is a central peak at $\kappa\,x = 0$ and two half-cut peaks at the boundaries 
$\kappa \, x=\pm \pi$.

Thus, we have shown how to obtain sub-half-wavelength localization through 
monitoring the probe absorption at a particular frequency.  We note that the 
atom is to be prepared in its ground state to start with as opposed to the 
schemes based on the observation of the  spontaneous emission spectrum 
(e.g., Ref.~\cite{GhafoorPRA2002}), where the atom needs to be prepared in 
its excited state. Thus the preparation stage is considerably simplified in 
our model. Moreover, as we need monitoring the probe absorption as opposed to 
spontaneous emission as in  Ref.~\cite{GhafoorPRA2002} 
we have distinct advantage to offer as the absorption measurements are 
straightforward to realize in an experiment compared to the measurement of 
spontaneous emission spectrum. In the following section we summarize our 
conclusions.

\section{CONCLUSIONS}
\label{sec:conclude}
We have studied an extended $\Lambda$--type scheme and have demonstrated appearance of phase dependence in the response of the atomic medium to a weak probe field.  We have further investigated application of the scheme to two situations one concerning the group velocity of the probe field and another allowing sub-wavelength localization of a moving atom.  

The model shows a wide range of tunability in the group velocity of the probe
field just by changing the phase of one of the control fields. 
The group velocity can also be switched from subluminal to superluminal
through a continuous change of the phase.
In contrast with most of the proposals where superluminality is accompanied with
considerable absorption, we can reduce the absorption of the probe pulse
substantially. The model imparts control of the propagation properties of the
probe pulse by controlling the intensities of the laser fields as well. In addition, the model imparts natural Doppler-free situation when all the fields are propagating in a collinear fashion. Moreover, we work within the EIT domain thus with sufficiently strong drive fields one does not have to worry about the absorption of any of the fields incident on the medium.

By taking one of the drive fields as a standing-wave field of a cavity we can use our model for sub-wavelength localization of the atom as it passes through the standing-wave.   Measurement of absorption of the probe field at a particular
frequency localizes the atom. We have shown that the precision of the  position measurement of the atom  depends upon the amplitudes and the relative phase of
the driving fields. The amplitude of standing-wave driving field when increased leads to  line narrowing in the probe absorption, thus giving increased precision in the position measurement. Whereas the relative phase of the
fields has important role in reducing the number of localization
peaks leading to sub-half-wavelength localization, namely confinement of the localization peaks to one of the half-wavelength regions of the cavity field.  
As the method is based on the measurement of the probe absorption, it has two distinct advantages compared to the similar methods based on the observation of the spontaneous emission spectrum. Absorption measurements are much easier to perform in a laboratory compared to monitoring of spontaneous emission spectrum. Moreover, we do not require the atoms to be prepared in their excited states, in fact they are prepared in their natural ground state.  Thus the preparation stage is fairly straightforward. 

\acknowledgments
Part of this work, done by KTK, was carried out 
at the Jet Propulsion Laboratory under 
a contract with the National Aeronautics and Space Administration (NASA). 
KTK acknowledges support from the National Research Council and
NASA, Code Y.  MSZ acknowledges  support of the Air
Force Office of Scientific Research,
DARPA-QuIST, TAMU
Telecommunication and Informatics Task Force (TITF) Initiative,
and the Office of
Naval Research.


\begin{thebibliography}{99}
\bibitem{1} M.~O. Scully, S.-Y. Zhu, and A. Gavrielides, ``Degenerate quantum-beat laser: Lasing without inversion and inversion without lasing,''  {\it Phys. Rev. Lett.} 
{\bf 62}, pp. 2813--2816, 1989; O. Kocharovaskaya, and P. Mandel, ``Amplification without inversion: The double- Lambda scheme,'' {\it Phys. Rev. A}
{\bf 42}, pp. 523--535, 1990; L. M. Narducci, {\it et al.}, ``A simple model of a laser without inversion,'' 
{\it Opt. Commun.} {\bf 81}, pp. 379--381, 1991. 

\bibitem{2} G. Alzetta, A. Gozzini, L. Moi and G. Orriols, ``Experimental method for observation of RF transitions and laser beat resonances in oriented Na vapor,''{\it 
Nuovo Cimento Soc. Ital. Fis.} {\bf B 36}, pp. 5--20, 1976. 

\bibitem{3} M.~O. Scully, ``Enhancement of the index of refraction via quantum coherence,'' {\it Phys. Rev. Lett.} {\bf 67}, pp. 1855--1858, 1991.  

\bibitem{4} S.~E. Harris, J.~E. field, and A.~Imomoglu, ``Nonlinear optical processes using electromagnetically induced transparency,''
{\it Phys. Rev. Lett.} {\bf 64}, pp. 1107--1110, 1990; 
K.~J. Boller, A. Imomoglu, and S.~E. Harris, ``Observation of electromagnetically induced transparency in collisionally broadened lead vapor,`` {\it ibid}  {\bf 67}, pp. 3062--3065, 1991. 

\bibitem{5} L.~V.  Hau, S.~E. Harris, Z.~Dutton and C.~H. Behroozi, ``Light speed reduction to 17 meters per second in an ultracold atomic gas,'' {\it Nature
(London)} {\bf  397}, pp. 594--598, 1999; M.~M. Kash et al.,  ``Ultraslow Group Velocity and Enhanced Nonlinear Optical Effects in a Coherently Driven Hot Atomic Gas,'' {\it Phys. Rev. Lett.} {\bf 82}, pp. 5229--5232, 1999.

\bibitem{wang}
L. J. Wang, A. Kuzmich, and A. Dogariu, ``Gain-assisted superluminal light propagation,'' {\it Nature (London)}, {\bf 406}, pp. 277--279 , 2000.


\bibitem{storage} D.~F. Phillips, M. Fleischhauer, A. Mair, R.~L. Wasworth, 
and M.~D. Lakin, ``Storage of Light in Atomic Vapor,'' {\it Phys. Rev. Lett.} {\bf 86}, pp. 783--786, 2001;  A. B. Matsko, ``Nonadiabatic approach to quantum optical information storage,'' {\it et al.} {\it Phys. Rev. A} {\bf 64}, Art. no. 043809, 2001.


\bibitem{6} S.-Y. Zhu, M.~O. Scully, ``Spectral Line Elimination and Spontaneous Emission Cancellation via Quantum Interference,'' {\it Phys. Rev. Lett.} {\bf 76}, pp. 388--391, 1996; P. Zhou, and S. Swain, ``Ultranarrow Spectral Lines via Quantum Interference,'' {\it Phys. Rev. Lett.} {\bf 77}, pp. 3995--3998, 1996;  E. Paspalakis, C.~H. Keitel, and P. L. Knight, ``Fluorescence control through multiple interference mechanisms,'' {\it Phys. Rev. A} {\bf 58}, pp. 4868--4877 1998; H. Lee, P. Polynkin, M.~O. Scully, and S.-Y. Zhu, ``Quenching of spontaneous emission via quantum interference,'' {\it Phys. Rev. A}  {\bf 55}, pp. 4454--4465 (1997); K. T. Kapale, M. O. Scully, S.-Y. Zhu, and M. S. Zubairy,  ``Quenching of spontaneous emission through interference of incoherent pump processes,'' {\it Phys. Rev. A} {\bf 67}, Art. no. 023804 (2003).

\bibitem{7} G.~S. Agrawal, {\it Quantum Statistical Theories of Spontaneous Emission and their Relation to other Approaches}, Edited by G. Hohler {\it et al.} Springer Tracts in Modern Physics Vol. {\bf 70}, Springer, Berlin, 1974. 

\bibitem{8} S.-Y. Zhu, L.~M. Narducci, and M.~O. Scully, ``Quantum-mechanical interference effects in the spontaneous-emission spectrum of a driven atom,'' {\it Phys. Rev. A} {\bf 52}, pp. 4791--4802, 1995; A. H. Toor, Shi-Yao Zhu, and M. S. Zubairy, `` Quantum interference in the spectrum of a driven atom: Effects of pumping and phase fluctuations,'' {\it Phys. Rev. A} {\bf 52}, pp. 4803--4811, 1995. 

\bibitem{Zoller96} H. Holland, S. Marksteiner, P. Marte and P. Zoller, ``Measurement Induced Localization from Spontaneous Decay,''
{\it Phys. Rev. Lett.} {\bf 76}, pp. 3683--3686, 1996.

\bibitem{Herkomer97} A. M. Herkommer, W. P. Schleich, and M. S. Zubairy, ``Autler-Townes microscopy on a single atom,'' {\it J. Mod.
Opt.} {\bf 44}, pp. 2507--2513, 1997.

\bibitem{QamarPRA2000} S. Qamar, S.-Y. Zhu, and M. S. Zubairy, ``Atom localization via resonance fluorescence,''
{\it Phys. Rev. A.} {\bf 61}, Art. no. 063806, 2000.

\bibitem{QamarOC2000} S. Qamar, S.-Y. Zhu, and M. S. Zubairy, ``Precision localization of single atom using Autler-Townes microscopy,'' {\it Opt.
Commun.} {\bf 176}, pp. 409--416, 2000.

\bibitem{GhafoorPRA2000} F. Ghafoor, S.-Y. Zhu, and M. S. Zubairy, ``Amplitude and phase control of spontaneous emission,''
{\it Phys. Rev. A.} {\bf 62}, Art. no. 013811, 2000.

\bibitem{GhafoorPRA2002} F. Ghafoor, S. Qamar, M. S. Zubairy, ``Atom localization via phase and amplitude control of the driving field,''
{\it Phys. Rev. A} {\bf 65}, Art. no. 043819, 2002.

\bibitem{Knight2001} E. Paspalakis, and P. L. Knight, ``Localizing an atom via quantum interference,'' {\it Phys. Rev. A.}  {\bf 63}, Art. no. 065802, 2001.

\bibitem{9} E. Paspalakis and P.~L. Knight, ``Phase Control of Spontaneous Emission,'' {\it Phys. Rev. Lett.} {\bf 81}, 
pp. 293--296, 1998.  

\bibitem{11} A. M. G. Martinez, P. R. Herczfeld, C. Samuels, L. M. Narducci, 
and C. H. Keitel, ``Quantum interference effects in spontaneous atomic emission: Dependence of the resonance fluorescence spectrum on the phase of the driving field,'' {\it Phys. Rev. A} {\bf 55}, pp. 4483--4491, 1997. 

\bibitem{arbiv}
D. Bortman-Arbiv, A. D. Wilson-Gordon, and H. Friedmann, ``Phase control of group velocity: From subluminal to superluminal light propagation,'' {\it Phys. 
Rev. A} {\bf 63}, Art. no. 043818, 2001.

\bibitem{SahraiGroupVel} M. Sahrai, H. Tajalli, K. T. Kapale, and M. S. Zubairy, ``Tunable phase control for subluminal to superluminal light propagation,''
{\it Phys. Rev. A} {\bf 70}, Art. no. 023813, 2004.

\bibitem{SahraiAL} M. Sahrai, H. Tajalli, K. T. Kapale, and M. S. Zubairy, ``Sub-wavelength atom-localization via amplitude and phase control of
the absorption spectrum,'' submitted to {\it Phys. Rev. A.}

\bibitem{wilson}
 A. D. Wilson-Gordon and H. Friedmann, ``Positive and negative dispersion in a three-level A system driven by a single pump,'' {\it J. Mod. Opt.} {\bf 49}, pp. 125--139, 2002.

\bibitem{agarwal}
G. S. Agarwal, T. N. Dey, and S. Menon,  ``Knob for changing light propagation from subluminal to superluminal,'' {\it Phys. Rev. A} {\bf 64}, Art. no. 053809, 2001.

\bibitem{goren}
C. Goren, A. D. Wilson-Gordon, M. Rosenbluh, and H. Friedmann, ``Switching from positive to negative dispersion in transparent degenerate and near-degenerate systems ,''
{\it Phys. Rev. A} {\bf 68}, Art. no. 043818, 2003.

\bibitem{matsko}
A. B. Matsko, O. Kocharovskaya, Y. Rostovtsev, G. R. Welch, A. S. Zibrov, and M. O. Scully, ``Slow, ultraslow, stored, and frozen light,''
{\it Adv. At. Mol. Opt. Phys.} {\bf 46}, pp. 191--242, 2001. (See pg. 208)

\end{thebibliography}
\end{document}